\documentclass[12pt]{amsart}
\usepackage{amssymb}
\usepackage{txfonts}
\usepackage{amsfonts}
\usepackage{mathrsfs}
 \theoremstyle{definition}
 
 \theoremstyle{remark}

 \numberwithin{equation}{section}

\begin{document}
\title[Witnesses detect  entangled states
simultaneously] {When different  entanglement witnesses detect
entangled states simultaneously}

\author{Jinchuan Hou}
\address{Department of Mathematics, Taiyuan
University of Technology,
 Taiyuan 030024, China; Department of Mathematics, Shanxi University,
Taiyuan, 030006, China } \email{jinchuanhou@yahoo.com.cn}

 \author{Yu Guo}
\address{Department of Mathematics, Shanxi University,
Taiyuan, 030006, China; Department of Mathematics, Shanxi Datong
University, Datong, 037009, China.}
 \email{guoyu3@yahoo.com.cn}

\thanks{This work is partially supported by National Natural Science Foundation of
China (10771157) and Research Fund of Shanxi for Returned Scholars
(2007-38).}

\begin{abstract}

The question under what conditions different witnesses may detect
some entangled states simultaneously is answered for both finite-
and infinite-dimensional bipartite systems. Finite many different
witnesses can detect some entangled states simultaneously if and
only if their convex combinations are still witnesses; they can not
detect any entangled state simultaneously if and only if the set of
their convex combinations contains  a positive operator. For two
witnesses $W_1$ and $W_2$, some more can be said:  (1) $W_1$ and
$W_2$ can detect the same set of entangled states if and only if
they are linearly dependent; (2) $W_2$ can detect more entangled
states than
 that $W_1$ can if and only if $W_1$ is a linear combination of $W_2$ and
a positive operator. As an  application, some characterizations of
the optimal witnesses are given and some structure properties of the
decomposable optimal witnesses are presented.

\end{abstract}

\maketitle

\section{Introduction}

The challenging question of characterizing the quantum entangled
states has attracted much attention in recent years. However,
despite remarkable progress in this field, there is no general
qualitative and quantitative characterizing of entanglement
\cite{CW,GH,GO1,GO2,HM1,HM2,HM3,HP1,LKH,NK,PA}.

Recall that, a bipartite quantum state (or density operator) in an
 bipartite system is a positive trace one
operator $\rho$ (i.e. $\rho\geq0$ and ${\rm Tr}(\rho)=1$) acting on
a complex tensor product Hilbert space $H_1\otimes H_2$ which
describing the bipartite quantum system, where $H_1$ and $H_2$ are
complex separable Hilbert spaces describing the corresponding
subsystems (we also say a unit vector in the corresponding Hilbert
space is a pure state). If both $H_1$ and $H_2$ are
finite-dimensional, then the composite system is a
finite-dimensional system; if at least one of $H_1$ and $H_2$ is
infinite-dimensional, then the composite system is an
infinite-dimensional system. By $\mathcal{S}^{(1)}={\mathcal
S}(H_1)$, $\mathcal{S}^{(2)}={\mathcal S}(H_2)$ and
$\mathcal{S}={\mathcal S}(H_1\otimes H_2)$ we denote the sets of all
states on $H_1$, $H_2$ and $H_1\otimes H_2$, respectively. Not that,
when $\dim H_1\otimes H_2=\infty$, we have ${\mathcal
S}\subset{\mathcal T}(H_1\otimes H_2)$, the Banach space of all
trace-class operators on $H_1\otimes H_2$ with trace norm
$\|\cdot\|_{\rm Tr}$. A state $\rho\in\mathcal{S}$ is said to be
\emph{separable} if it is a trace-norm limit of the states of the
form
$$\rho=\sum\limits_ip_i\rho_i^{(1)}\otimes\rho_i^{(2)},$$ where
 $\rho_i^{(1)}$ and
$\rho_i^{(2)}$ are pure states in $\mathcal{S}^{(1)}$ and
$\mathcal{S}^{(2)}$, respectively, $\sum\limits_ip_i=1$, $p_i\geq0$.
Otherwise, $\rho$ is said to be \emph{entangled} (or inseparable).
The set of all separable states will be denoted by ${\mathcal
S}_{\rm sep}(H_1\otimes H_2)$.

Among the multitudinous criteria for deciding whether a given state
is entangled or not, the well known one is the \emph{entanglement
witness criterion} \cite{HM1}. This criterion provides a sufficient
and necessary condition for separability of a given state in a
bipartite quantum system. It is shown that
 \cite{HM1}, a given state is separable if and only if there exists
at least one entanglement witness detecting it. A self-adjoint
operator (also called hermitian operator some times) $W$ acting on
$H_1\otimes H_2$ is called an \emph{entanglement witness} (or
\emph{witness} for short) if ${\rm Tr}(W\sigma)\geq0$ for all
separable sates $\sigma\in{\mathcal S}_{\rm sep}$ and ${\rm
Tr}(W\rho)<0$ for at least one entangled state $\rho$ (in this case,
we say that $\rho$ is detected by $W$, or, equivalently, $W$ is a
witness for $\rho$).

Although any entangled state can be detected by some specific choice
of witness, there is no universal witness, i.e., there is no witness
which can detect all entangled states. From the entanglement witness
criterion,   the task is reduced to find out all witnesses. However,
constructing the witnesses for an entangled state is a hard task,
and the determination of witnesses for all entangled states is a
NP-hard problem \cite{BV2}.

Witnesses not only can be used to detect any entangled states, but
also are directly measurable quantities.  This makes the
entanglement witnesses one of the main methods to detect
entanglement experimentally and a very useful tool for analyzing
entanglement in experiment. So, it is important to know more about
the features of the witnesses. Concerning this topic, much work has
been done for finite-dimensional systems (for example, ref.
\cite{LK, WH}). However, few results are known for
infinite-dimensional systems. Generally, the structure of witnesses
for infinite-dimensional systems  are complicated. However, it was
proved in \cite{QX} that, for any entangled state, a witness  can be
chosen so that it has a simple form of ``nonnegative constant times
identity + a self-adjoint operator of finite rank''. This kind of
witnesses are Fradholm operator of index 0 with the spectrum
consisting of finite many eigenvalues and hence are easily handled.
 The goal of the present paper is to solve the question when deferent  witnesses
 can detect some entangled states simultaneously for mainly infinite-dimensional systems.

For simplicity, we introduce some notations. Let $H_1,H_2$ be
complex Hilbert spaces and let ${\mathcal W}= {\mathcal
W}(H_1\otimes H_2)$ be the set of all entanglement witnesses of the
system $H_1\otimes H_2$, i.e.,
$$\begin{array}{rl}\mathcal{W}=& {\mathcal W}(H_1\otimes H_2)\\ =&
\{W: W\in{\mathcal B}(H_1\otimes H_2), W^\dagger=W,\\ & {\rm
Tr}(W\sigma)\geq0\ \mbox{\rm for all }\sigma\in\mathcal{S}_{\rm sep}
\ {\rm and}\ W\ \mbox{\rm is not positive}\}.\end{array}$$  For
$W\in\mathcal{W}$ and $\Gamma\subset \mathcal{W}$, define
$$\mathcal{D}_W=\{\rho:\rho\in{\mathcal S}, {\rm Tr}(W\rho)<0\}$$ and
$\mathcal{D}_{\Gamma}=\bigcap\limits_{W\in\Gamma}\mathcal{D}_W$.
Then $\mathcal{D}_W$ and $\mathcal{D}_{\Gamma}$ are convex sets.
Thus the witnesses in $\Gamma$ can detect some entangled states
simultaneously if and only if ${\mathcal D}_\Gamma\not=\emptyset$.

For $W_1,W_2\in{\mathcal W}$, if $\mathcal{D}_{W_2}\subset
\mathcal{D}_{W_1}$, we say that $W_1$ is \emph{finer} than $W_2$,
denoted by
$$W_2\prec W_1.$$ We call $W_1$ is an \emph{optimal} witness if there
exists no other witness  finer than $W_1$. Then $\mathcal{W}$
becomes a poset with respect to the partial order $``\prec"$.
Generally, for two given witnesses $W_1$ and $W_2$, there are three
different situations that may happen: (i)
$\mathcal{D}_{W_1}\subseteq\mathcal{D}_{W_2}$, i.e., $W_1\prec W_2$,
in particular, $\mathcal{D}_{W_1}=\mathcal{D}_{W_2}$; (ii)
$\mathcal{D}_{W_1}\cap\mathcal{D}_{W_2}\not=\emptyset$  and
$\mathcal{D}_{W_i}\nsubseteq\mathcal{D}_{W_j}$, $i$, $j=1, 2$; (iii)
$\mathcal{D}_{W_1}\cap\mathcal{D}_{W_2}=\emptyset$. Thus $W_1$ and
$W_2$ can detect a state simultaneously if and only if (i) or (ii)
holds.

For the finite-dimensional case, the relations (i)-(iii) above are
studied in \cite{LK,WH}. Suppose that ${\rm Tr}(W_1)={\rm Tr}(W_2)$,
then the following conclusions are true: (i)
$\mathcal{D}_{W_1}\subseteq\mathcal{D}_{W_2}$ if and only if
$W_1=(1-\varepsilon)W_2+\varepsilon D$ for some $D\geq0$ and
$0\leq\varepsilon<1$; in particular,
$\mathcal{D}_{W_1}=\mathcal{D}_{W_2}$ if and only if $W_1=W_2$
\cite{LK}; (ii) if there are no inclusion relations between
$\mathcal{D}_{W_1}$ and $\mathcal{D}_{W_2}$, then
$\mathcal{D}_{W_1}\cap\mathcal{D}_{W_2}\neq\emptyset$ if and only if
$W=\epsilon W_1+(1-\epsilon)W_2$ is not positive for any
$0\leq\epsilon\leq1$ \cite{WH}. However, there are some mistakes in
the proof of \cite{WH}. The main purpose of the present paper is to
show that the similar results holds for infinite-dimensional systems
and to correct the mistakes appeared in \cite{WH}. Note that, the
condition ${\rm Tr}(W_1)={\rm Tr}(W_2)$ makes no sense in general
for infinite-dimensional case.

This paper is organized as follows. In Section 2, we propose a
sufficient and necessary condition for any two given general witness
$W_1$ and $W_2$ to satisfy  $W_1\prec W_2$. Let $H_1$, $H_2$ be
complex Hilbert spaces. Assume that
$W_1$,$W_2\in\mathcal{W}(H_1\otimes H_2)$. We show that, (1)
$W_1\prec W_2$ if and only if $W_1=aW_2+D$ for some
 operator $D\geq 0$ and some scalar $a> 0$;
(2)   $\mathcal{D}_{W_1}=\mathcal{D}_{W_2}$ if and only if there
exists a positive number $a>0$ such that $W_1=a W_2$. Then these
results are applied   in Section 3 to obtain a sufficient and
necessary condition for a witness to be optimal. We show that
$W\in\mathcal{W}(H_1\otimes H_2)$  is optimal if and only if for any
nonzero operator $D\geq 0$ and scalar $a>0$,
$W^\prime=aW-D\notin\mathcal{W}(H_1\otimes H_2)$. Some structure
properties of the optimal decomposable witnesses are also presented.
 In Section 4, we discuss the question when finite many witnesses
 can detect a common
entangled state. We show that
$\cap_{k=1}^n\mathcal{D}_{W_k}\not=\emptyset$ if and only if every
combination of $W_1, \ldots , W_n$ is still a witness;
$\cap_{k=1}^n\mathcal{D}_{W_k}=\emptyset$ if and only if there
exists at least one convex combination $W$ of $W_1, \ldots , W_n$
such that $W\geq 0$.

Throughout this paper, we call an operator $A\in\mathcal{B}(H)$ is
positive, if $\langle x|A|x\rangle\geq0$ for all $|x\rangle\in H$.
$\|\cdot\|_{\rm Tr}$ denotes the trace norm, and $\|\cdot\|_2$
denotes the Hilbert-Schmidt norm. For a operator $A$, $A^T$ stands
for the transposition of $A$ with respect some given orthonormal
basis. By $A^{T_2}$ we denote the partial transposition of $A$ with
respect to the second subsystem $H_2$, i.e., $A^{T_2}=(I_1\otimes
\tau)A$, where $\tau $ is the transpose operation. \if As usual, we
denote by $\rho_A$ and $\rho_B$ the reduced states of $\rho$ with
respect to the corresponding subsystem, that is $\rho_A={\rm
Tr_B}(\rho)\in\mathcal{S}(H)$ and $\rho_B={\rm
Tr_A}(\rho)\in\mathcal{S}(K)$.\fi $\mathcal{T}(H_1\otimes H_2)$
denotes the set of all trace class operators in
$\mathcal{B}(H_1\otimes H_2)$ while $\mathcal{T}^+ (H_1\otimes H_2)$
stands for the set of all positive elements in
$\mathcal{T}(H_1\otimes H_2)$.

\section{ Witnesses with the  finer relation between them  }

In this section, we mainly highlight the {\it finer} relation
between two given general witnesses of an {\it infinite-dimensional}
bipartite system.

For finite-dimensional bipartite quantum system, it is known that if
$W_1$, $W_2\in\mathcal{W}$  with ${\rm Tr}(W_1)={\rm Tr}(W_2)$, then
$\mathcal{D}_{W_1}\subseteq\mathcal{D}_{W_2}$ if and only if
$W_1=(1-\varepsilon)W_2+\varepsilon D$ for some $D\geq0$ and
$0\leq\varepsilon<1$; $\mathcal{D}_{W_1}=\mathcal{D}_{W_2}$ if and
only if $W_1=W_2$ \cite{LK}.
 Since the condition ${\rm
Tr}(W_1)={\rm Tr}(W_2)$ makes no sense in general for
infinite-dimensional case, we have to consider the question without
the trace-equal assumption.

The following is the main result  in this section which answers the
question when ${\mathcal D}_{W_1}\cap{\mathcal D}_{W_2}={\mathcal
D}_{W_1}$ for both infinite-dimensional systems and
finite-dimensional cases.

{\bf Theorem 2.1.} \ {\it Let $H_1$, $H_2$ be complex Hilbert
spaces. Assume that $W_1$,$W_2\in\mathcal{W}(H_1\otimes H_2)$. Then
}

(1) {\it   $W_1\prec W_2$ if and only if $W_1=aW_2+D$ for some
 operator $D\geq 0$ and some scalar $a> 0$.}

(2) {\it $\mathcal{D}_{W_1}=\mathcal{D}_{W_2}$ if and only if there
exists a positive number $a>0$ such that $W_1=a W_2$.}

To prove Theorem 2.1, we need several  lemmas.

We first generalize a useful result in \cite{LK} to
infinite-dimensional case, which asserts that the restriction of any
entanglement witness as a linear functional to the convex set
consisting of separable states is nonzero.

{\bf Lemma 2.2.} \ {\it  Let $H_1$, $H_2$ be complex Hilbert spaces.
For any $W\in{\mathcal W}(H_1\otimes H_2)$, there is a separable
pure state $\sigma\in{\mathcal S}_{\rm sep}(H_1\otimes H_2)$ such
that ${\rm Tr}(W\sigma)>0$.}

{\bf Proof.} \ Let $\{|i\rangle\}$ and $\{|j\rangle\}$ be any
orthonormal bases of $H_1$ and $H_2$, respectively. Then,
$\{|i\rangle|j\rangle\}$ is an orthonormal basis of $H_1\otimes
H_2$. It turns out  $\langle i|\langle j|W|i\rangle|j\rangle\geq0$
since $\langle i|\langle j|W|i\rangle|j\rangle={\rm
Tr}(W|i\rangle\langle i|\otimes |j\rangle\langle j|)\geq0$ for any
$i$, $j$.

 To prove the lemma, it is suffice to show that
there exist orthonormal bases $\{|i\rangle\}$ and $\{|j\rangle\}$
such that ${\rm Tr}(W|i\rangle\langle i|\otimes |j\rangle\langle
j|)\neq 0$ for some $i,j$. To get a contradiction, assume that this
is not true.  Then
$$\langle\psi_1|\langle\psi_2|W|\psi_1\rangle|\psi_2\rangle=0$$ for
all product vectors $|\psi_1\rangle|\psi_2\rangle\in H_1\otimes
H_2$. For any pure state $|\psi\rangle\in H_1\otimes H_2$, let
$|\psi\rangle=\sum\limits_{k=1}^n\lambda_k|k\rangle|k^{\prime}\rangle$
be the Schmidt decomposition of $|\psi\rangle$, where $\lambda_k>0$,
$\sum _{k=1}^n\lambda_k^2=1$ and $\{|k\rangle\}_{k=1}^n,
\{|k^\prime\rangle\}_{k^\prime=1}^n$ are orthonormal sets
respectively in $H_1,H_2$, here $n$ is called the Schmidt number of
$|\psi\rangle$. Then,
$$\begin{array}{rl}\rho=&|\psi\rangle\langle\psi|\\
=&(\sum\limits_k\lambda_k|k\rangle|k^{\prime}\rangle)
(\sum\limits_l\lambda_l\langle l |\langle l^{\prime}|)\\
=&\sum\limits_{k,l}\lambda_k\lambda_l|k\rangle\langle
l|\otimes|k^{\prime}\rangle\langle l^{\prime}|\\
=&\sum\limits_{k=l}\lambda_k^2|k\rangle\langle
k|\otimes|k^{\prime}\rangle\langle k^{\prime}|+\sum\limits_{k<
l}\lambda_k\lambda_l(|k\rangle\langle
l|\otimes|k^{\prime}\rangle\langle l^{\prime}|+|l\rangle\langle
k|\otimes|l^{\prime}\rangle\langle k^{\prime}|).\end{array}$$ For
given pair $(k$, $l)$ with $k\not= l$, define
$|\psi_{k,l}\rangle=\frac{1}{\sqrt{2}}(|k\rangle|k^{\prime}\rangle
+|l\rangle|l^{\prime}\rangle)$. We have $$|k\rangle\langle
l|\otimes|k^{\prime}\rangle\langle l^{\prime}|+|l\rangle\langle
k|\otimes|l^{\prime}\rangle\langle
k^{\prime}|=2|\psi_{k,l}\rangle\langle\psi_{k,l}|-|k\rangle\langle
k|\otimes|k^{\prime}\rangle\langle k^{\prime}|-|l\rangle\langle
l|\otimes|l^{\prime}\rangle\langle l^{\prime}|.$$ This indicates
that, if $n<\infty$, then $\langle\psi|W|\psi\rangle =0$. As the set
of all unit vectors with the finite Schmidt number is dense in the
set of all unit vectors in $H_1\otimes H_2$, we see that
$\langle\psi|W|\psi\rangle =0$ holds for all unit vector
$|\psi\rangle$ and hence $W=0$, a contradiction. \hfill{$\square$}

 Analogues to the
finite-dimensional case \cite{LK}, the following lemma is obvious.

{\bf Lemma 2.3.} \ {\it  Let $H_1$, $H_2$ be complex Hilbert spaces.
For a given $W\in\mathcal{W}(H_1\otimes H_2)$, if
$\rho\in\mathcal{D}_W$ and $\varrho_W\in\mathcal{T}^+(H_1\otimes
H_2)$ satisfying ${\rm Tr}(W\varrho_W)=0$, then
$(\rho+\varrho_W)/{\rm Tr}(\rho+\varrho_W)\in \mathcal{D}_W$.}

The next lemma is crucial for our purpose. Its statement as well as
its proof are quite different from the counterpart lemma in
\cite{LK} for finite-dimensional case.

{\bf Lemma 2.4.} \ {\it Let $H_1$, $H_2$ be complex Hilbert spaces
and $W_1$, $W_2\in\mathcal{W}(H_1\otimes H_2)$. Assume that
$W_1\prec W_2$ and let
$$\lambda:=\inf\limits_{\rho_1\in\mathcal{D}_{W_1}}\frac{|{\rm
Tr}(W_2\rho_1)|}{|{\rm Tr}(W_1\rho_1)|}.$$ Then  the following
statements are true:}

(1) {\it If $\rho\in\mathcal{S}(H_1\otimes H_2)$ satisfies ${\rm
Tr}(W_1\rho)=0$, then ${\rm Tr}(W_2\rho)\leq0$;}

(2) {\it $\lambda >0$. \if false such that, if $\rho\in\mathcal{S}$
satisfies ${\rm Tr}(W_1\rho)<0$, then ${\rm Tr}(W_2\rho)\leq
\alpha{\rm Tr}(W_1\rho)$;\fi }

(3) {\it If $\rho\in\mathcal{S}(H_1\otimes H_2)$ satisfies ${\rm
Tr}(W_1\rho)>0$, then ${\rm Tr}(W_2\rho)\leq\lambda{\rm
Tr}(W_1\rho)$.}

{\bf Proof.} \ (1) Let us assume, to reach a contradiction, that
${\rm Tr}(W_2\rho)>0$. Then, for any $\rho_1\in\mathcal{D}_{W_1}$
and $a\geq0$, we have
$\rho(a)=(\rho_1+a\rho)/(1+a)\in\mathcal{D}_{W_1}$. On the other
hand, there exists a positive number $a_0$ such that ${\rm
Tr}(W_2\rho(a))>0$ holds for all $a\geq a_0$, which is impossible
since it leads to $\rho(a)\notin\mathcal{D}_{W_2}$.

(2) Assume that, on the contrary, $\lambda =0$.  Then,  there exists
a sequence $\{\rho_n\}\subset{\mathcal D}_{W_1}$ such that
$$\varepsilon_n=\frac{{\rm Tr}(W_2\rho_n)}{{\rm
Tr}(W_1\rho_n)}\rightarrow 0\ \mbox{as}\
n\rightarrow\infty.\eqno(2.1)$$

Note that there exists $\sigma\in{\mathcal S}_{\rm sep}={\mathcal
S}_{\rm sep}(H_1\otimes H_2) $ such that both ${\rm Tr}(W_1\sigma)$
and ${\rm Tr}(W_2\sigma)$ are nonzero. If not, then for any
$\sigma\in{\mathcal S}_{\rm sep}$, either ${\rm Tr}(W_1\sigma)=0$ or
${\rm Tr}(W_2\sigma)=0$. Thus, by Lemma 2.2, there exist
$\sigma_1,\sigma_2\in{\mathcal S}_{\rm sep} $ so that ${\rm
Tr}(W_1\sigma_1)=t>0$, ${\rm Tr}(W_1\sigma_2)=0$, ${\rm
Tr}(W_2\sigma_1)=0$ and ${\rm Tr}(W_2\sigma_2)=s>0$. Let
$\sigma=\frac{s}{t+s}\sigma_1+\frac{t}{t+s}\sigma_2\in{\mathcal
S}_{\rm sep} $. Then ${\rm Tr}(W_1\sigma)={\rm
Tr}(W_2\sigma)=\frac{ts}{t+s}\not=0$, contradicting to the
assumption.

Now we can take  $\sigma\in{\mathcal S}_{\rm sep}$ so that both
${\rm Tr}(W_1\sigma)$ and ${\rm Tr}(W_2\sigma)$ are nonzero. Let
$$\tilde{\rho}_n=\frac{1}{1-\frac{{\rm Tr}(W_1\rho_n)}{{\rm
Tr}(W_1\sigma)}}(\rho_n-\frac{{\rm Tr}(W_1\rho_n)}{{\rm
Tr}(W_1\sigma)}\sigma)\in{\mathcal S}$$ with $\rho_n$ satisfying
Eq.(2.1). Then ${\rm Tr}(W_1\tilde{\rho}_n)=0$ and by (1), we have
${\rm Tr}(W_2\tilde{\rho}_n)\leq 0$ for every $n$. However,
$$\begin{array}{rl}{\rm
Tr}(W_2\tilde{\rho}_n)=&\frac{1}{1-\frac{{\rm Tr}(W_1\rho_n)}{{\rm
Tr}(W_1\sigma)}}({\rm Tr}(W_2\rho_n)-\frac{{\rm Tr}(W_1\rho_n)}{{\rm
Tr}(W_1\sigma)}{\rm Tr}(W_2\sigma))\\=&\frac{1}{1-\frac{{\rm
Tr}(W_1\rho_n)}{{\rm Tr}(W_1\sigma)}}(\varepsilon_n-\frac{{\rm
Tr}(W_2\sigma)}{{\rm Tr}(W_1\sigma)}){\rm
Tr}(W_1\rho_n)\end{array}$$ and $\varepsilon_n\rightarrow 0$, which
implies that for sufficient large $n$, we have
$\varepsilon_n-\frac{{\rm Tr}(W_2\sigma)}{{\rm Tr}(W_1\sigma)}<0$
and hence ${\rm Tr}(W_2\tilde{\rho}_n)>0$, a contradiction. This
completes the proof of (2).

(3) Assume that ${\rm Tr}(W_1\rho)>0$. Take $\rho_1\in\mathcal{
D}_{W_1}$ and let $\tilde{\rho}=\frac{1}{{\rm Tr}(W_1\rho)-{\rm
Tr}(W_1\rho_1)}[{\rm Tr}(W_1\rho)\rho_1-{\rm Tr}(W_1\rho_1)\rho]$.
Then we have ${\rm Tr}(W_1\tilde{\rho})=0$. By (1), we obtain that
${\rm Tr}(W_2\tilde{\rho})\leq0$. Thus we have ${\rm
Tr}(W_1\rho){\rm Tr}(W_2\rho_1)\leq{\rm Tr}( W_1\rho_1){\rm
Tr}(W_2\rho)$. It follows that
$$\frac{{\rm Tr}(W_2\rho)}{{\rm Tr}(W_1\rho)}\leq\frac{|{\rm
Tr}(W_2\rho_1)|}{|{\rm Tr}(W_1\rho_1)|}.$$ Taking the infimum with
respect to $\rho_1\in \mathcal{D}_{W_1}$ on the right side of the
above equation, we get ${\rm Tr}(W_2\rho)\leq\lambda{\rm
Tr}(W_1\rho)$.\hfill{$\square$}

Now we are in a position to give our proof of Theorem 2.1.

{\bf Proof of Theorem 2.1.} \ (1) If $W_1=aW_2+D$ for some positive
operator $D$ and some scalar $a>0$, then for any $\rho\in{\mathcal
D}_{W_1}$, we have $a{\rm Tr}(W_2\rho)+{\rm Tr}(D\rho)={\rm
Tr}(W_1\rho)<0$, which implies that ${\rm Tr}(W_2\rho)<0$. Hence
${\mathcal D}_{W_1}\subseteq{\mathcal D}_{W_2}$. Conversely, assume
that ${\mathcal D}_{W_1}\subseteq{\mathcal D}_{W_2}$. Then, by Lemma
2.4,
$${\rm Tr}(W_2\rho)\leq \lambda {\rm Tr}(W_1\rho) \eqno{(2.2)}$$
holds for all $\rho\in{\mathcal S}$, where
$\lambda=\inf\limits_{\rho_1\in\mathcal{D}_{W_1}}\frac{|{\rm
Tr}(W_2\rho_1)|}{|{\rm Tr}(W_1\rho_1)|}>0$. This implies that
$D_1=\lambda W_1-W_2 \geq 0$ and hence, with $D=\lambda^{-1} D_1$,
$W_1=\lambda^{-1} W_2+D$, as desired.

(2) We only need to prove the `only if' part. Assume that ${\mathcal
D}_{W_1}={\mathcal D}_{W_2}$. Then, by the statement (1) just proved
above, there exist operators $D_i\geq 0$ and scalars $a_i>0$,
$i=1,2$, such that $W_1=a_1W_2+D_1$ and $W_2=a_2W_1+D_2$. It follows
that $W_1=a_1(a_2W_1+D_2)+D_1=a_1a_2W_1+a_1D_2+D_1$. Thus
$(1-a_1a_2)W_1=a_1D_2+D_1\geq 0$. Since $W_1\in{\mathcal W}$, we
must have $a_1a_2=1$. Hence $D_1=D_2=0$ and $W_2=a_2W_1$, completing
the proof.\hfill$\square$

\if false  Define $\lambda$ as in Lemma 2.5. Similarly, we define
$$\lambda^\prime:=\inf\limits_{\rho_2\in \mathcal{D}_{W_2}}\frac{|{\rm
Tr}(W_2\rho_2)|}{|{\rm Tr}(W_1\rho_2)|},$$ By Lemma 2.5, we know
that $\lambda^\prime\geq1$ as $W_2\prec W_1$. Thus,
$$\sup\limits_{\rho_2\in\mathcal{D}_{W_2}}\frac{|{\rm Tr}(W_2\rho_2)|}{|{\rm
Tr}(W_1\rho_2)|}=\sup\limits_{\rho_1\in\mathcal{D}_{W_1}}\frac{|{\rm
Tr}(W_2\rho_1)|}{|{\rm Tr}(W_1\rho_1)|}\leq1/\lambda^\prime\leq1.$$
On the other hand,
$\sup\limits_{\rho_1\in\mathcal{D}_{W_1}}\frac{|{\rm
Tr}(W_2\rho_1)|}{|{\rm Tr}(W_1\rho_1)|}\geq\lambda\geq1$. It follows
that $\lambda=1$, by Lemma 2.5. (4), we have $W_1=\delta W_2$ for
some $\delta>0$.\hfill{$\square$}

Next, we show a way of finding a witness which is finer than a given
one.

{\bf Theorem 2.6.} \ {\it If $W_1$, $W_2\in\mathcal{W}$, then
$W_1\prec W_2$ if and only if $W_1=aW_2+bD$ for some $D\geq0$ and
$a\geq0$, $b\geq0$.}

{\bf Proof.} \ The `if' part. For any $\rho\in\mathcal{D}_{W_1}$, we
have ${\rm Tr}(W_1\rho)<0$, that is $a{\rm Tr}(W_2\rho)+b{\rm
Tr}(D\rho)<0$. It is clear that ${\rm Tr}(W_2\rho)<0$, thus
$W_1\prec W_2$.

The `only if' part. We define $$\lambda:=\inf\limits_{\rho_1\in
\mathcal{D}_{W_1}}\frac{|{\rm Tr}(W_2\rho_1)|}{|{\rm
Tr}(W_1\rho_1)|},$$ then $\lambda\geq1$. If $\lambda=1$, then, by
Lemma 2.5, we know that $W_1=\delta W_2$ for some $\delta>0$, take
$a=\delta$ and $b=0$, that is $W_1=aW_2+bD$. If $\lambda>1$, let
$D=\frac{1}{\lambda-1}(\lambda W_1-W_2)$, it yields
$W_1=\frac{1}{\lambda}W_2+(1-\frac{1}{\lambda})D$. We show that
$D\geq0$ in the following. For any $\rho\in\mathcal{S}$, we have
${\rm Tr}(D\rho)=\frac{\lambda{\rm Tr}(W_1\rho)-{\rm
Tr}(W_2\rho)}{\lambda-1}\geq0$ since $\lambda{\rm
Tr}(W_1\rho)\geq{\rm Tr}(W_2\rho)$. It follows that ${\rm
Tr}(DA)\geq0$ for all $A\in\mathcal{C}_1^+(H_1\otimes H_2)$. In
fact, if $D$ is not positive, we may assume that $\lambda_0<0$ is a
eigenvalue of $D$ and $|\psi\rangle\in H_1\otimes H_2$ is the
associated unit eigenvector. Write
$\rho_\psi=|\psi\rangle\langle\psi|$, it follows that ${\rm
Tr}(D\rho_\psi)=\lambda_0<0$, a contradiction. Thus $D\geq0$ as
desired.\hfill{$\square$} \fi

\section{Optimization of entanglement witnesses}

In this section we discuss the optimization of entanglement
witnesses, especially for infinite-dimensional systems by applying
Theorem 2.1.

The following result states that a witness is optimal if and only if
any negative permutation if it will break the witness. For
finite-dimensional case, a similar result was obtained in \cite{LK}.

 {\bf Theorem 3.1.} \ {\it Let $H_1$, $H_2$ be
complex Hilbert spaces. Then $W\in\mathcal{W}(H_1\otimes H_2)$  is
optimal if and only if for any nonzero operator $D\geq 0$ and scalar
$a>0$, $W^\prime=aW-D\notin\mathcal{W}(H_1\otimes H_2)$.}

{\bf Proof.}\ To prove the `if' part,  assume, on the contrary, that
$W$ is not optimal, then $W\prec W^\prime$ for some
$W^\prime\in\mathcal{W}(H_1\otimes H_2)$ with $W$ and $W^\prime$ are
linearly independent. It follows from Theorem 2.1(1) that
$W=aW^\prime+ D$ for some $D\geq0$ and $a>0$, which reveals that
$W^\prime=\frac{1}{a}W-\frac{1}{a}D$.

To prove the `only if' part, assume that  $W$ is   optimal but there
exist   nonzero operator $D\geq0$, scalar $a>0$ so that $W^\prime
=aW-D \in{\mathcal W}(H_1\otimes H_2)$. Then
$W=\frac{1}{a}W^\prime+\frac{1}{a}D$ and $W^\prime$ is linearly
independent to $W$. But by Theorem 2.1, $W\prec W^\prime$, a
contradiction. \hfill{$\square$}

\if Similar to the finite-dimensional case \cite{LK}, the previous
theorem shows that an entanglement witness  is optimal if and only
if when we subtract any nonzero positive operator from it the
resulting operator is not an entanglement witness.  Notice that,
this result is not very practicable due to the fact that the
operator $D$ and the numbers $a$   are not easy to be determined.\fi

In the following, we discuss the condition for an entanglement
witness that it cannot subtract some positive operators. For
convenience, we define $$\mathcal{P}_{W}=\{\
|\psi\rangle|\phi\rangle\in H_1\otimes H_2:
\langle\psi|\langle\phi|W|\psi\rangle|\phi\rangle=0\ \}.\eqno(3.1)$$

{\bf Proposition 3.2.}\ {\it Let $H_1$, $H_2$ be complex Hilbert
spaces and $W\in{\mathcal W}(H_1\otimes H_2)$. Let $\mathcal{P}_{W}$
be as in Eq.(3.1). If $D\in{\mathcal B}(H_1\otimes H_2)$ is positive
and $D\mathcal{P}_{W}\neq\{0\}$, then $W-aD\notin{\mathcal
W}(H_1\otimes H_2)$ for any $a>0$.}

{\bf Proof.} \ If $D\mathcal{P}_W\neq\{0\}$, then there exists a
product vector $|\psi_0\rangle|\phi_0\rangle\in\mathcal{P}_W$ such
that
$$\langle\psi_0|\langle\phi_0|D|\psi_0\rangle|\phi_0\rangle>0.$$
Write
$\rho_0=|\psi_0\rangle\langle\psi_0|\otimes|\phi_0\rangle\langle\phi_0|$.
It is clear that ${\rm Tr}[(W-aD)\rho_0]=-a{\rm Tr}(D\rho_0)<0$,
which leads to $W-aD\notin{\mathcal W}(H_1\otimes H_2)$
 for
all $a>0$.\hfill{$\square$}

The following corollary is obvious.

{\bf Corollary 3.3.}\ {\it Let $H_1$, $H_2$ be complex Hilbert
spaces and $W\in{\mathcal W}(H_1\otimes H_2)$. Let $\mathcal{P}_{W}$
be as in Eq.(3.1). If $\mathcal{P}_W$ spans $H_1\otimes H_2$, then
$W$ is optimal.}

Next we give some structure properties of optimal decomposable
witnesses. Recall that a self-adjoint operator
$A\in\mathcal{B}(H_1\otimes H_2)$ is said to be \emph{decomposable}
if $$A=P+Q^{T_2}$$ for some operators $P\geq0$, $Q\geq0$, where
$Q^{T_2}$ denotes the partial transpose of $Q$ with respect to the
second subsystem $H_2$. Otherwise, $A$ is said to be
\emph{indecomposable}.  For example, in $n\times n$ system, the
Hermitian swap operator $V=\sum\limits_{i,j=0}^{n-1}|i\rangle\langle
j|\otimes|j\rangle\langle i|$ is a decomposable witness since: (1)
${\rm Tr}(V\sigma)\geq0$ for all separable pure states $\sigma$; (2)
$V$ has a negative eigenvalue -1; (3) $V=nQ^{T_2}$ with
$Q=|\psi\rangle\langle\psi|$ with
$|\psi\rangle=\frac{1}{\sqrt{n}}\sum\limits_{i=0}^{n-1}|i\rangle|i\rangle$
(ref. \cite{W}). The examples of indecomposable witnesses can be
found in \cite{CD,GM,QX}. It is easy to show that the decomposable
witnesses can not detect any  PPT entangled states (PPT stands for
\emph{positive partial transposition} as usual,  \cite{JB}).

By  applying Theorem 2.1, one can get a simple structure property of
optimal decomposable entanglement witnesses for both
finite-dimensional systems and infinite-dimensional systems.

 {\bf Theorem 3.4.} {\it Let $H_1$, $H_2$ be complex Hilbert
spaces and $W\in{\mathcal W}(H_1\otimes H_2)$ be a decomposable
entanglement witness. If $W$ is optimal, then $W=Q^{T_2}$ for some
$Q\geq0$, and $Q$ contains no product vectors in its range.}

{\bf Proof.}  Since $W$ is decomposable, so $W=P+Q^{T_2}$ for some
positive operators $P$, $Q$. Assume that $P\neq0$. As ${\rm
Tr}(Q^{T_2}\sigma)={\rm Tr}(Q\sigma^{T_2})\geq0$ for all
$\sigma\in\mathcal{S}_{\rm sep}$ and $W\in{\mathcal W}$, we must
have $Q^{T_2}\in\mathcal{W}$. Thus, by Theorem 2.1 (1), one sees
that $W\prec Q^{T_2}$, that is, $W$ is not optimal. Hence,   $W$ is
optimal implies that $P=0$ and $W=Q^{T_2}$. Moreover, the range of
$Q$ contains no product vectors. In fact, if
$|\psi\rangle|\phi\rangle\in R(Q)$ for some unit vectors
$|\psi\rangle\in H_1$ and $|\phi\rangle\in H_2$, then there exists a
vector $|\omega\rangle\in H_1\otimes H_2$ such that
$Q|\omega\rangle=|\psi\rangle\otimes|\phi\rangle$. Observe that
$Q(I-\lambda|\omega\rangle\langle\omega|)Q
=Q^2-\lambda|\psi\rangle\langle\psi|\otimes|\phi\rangle\langle\phi|\geq0$
if and only if $I-\lambda|\omega\rangle\langle\omega|\geq0$. It
turns out that, for any $0<\lambda<\||\omega\rangle\|^{-2}$ we have
$[Q-\lambda|\psi\rangle\langle\psi|\otimes|\phi\rangle\langle\phi|]^{T_2}\in\mathcal{W}$,
which implies that
$[Q-\lambda|\psi\rangle\langle\psi|\otimes|\phi\rangle\langle\phi|]^{T_2}$
is finer than $W$, contradicting to the optimality of
$W$.\hfill{$\square$}

Theorem 3.4 can be strengthened a little.

{\bf Theorem 3.5.}  {\it Let $H_1$, $H_2$ be complex Hilbert spaces
and $W\in{\mathcal W}(H_1\otimes H_2)$ be a decomposable
entanglement witness. If $W $ is optimal, then $W=Q^{T_2}$ for some
$Q\geq0$ and there exists no positive operator $A$ with
$R(A)\subseteq R(Q)$ such that $A^{T_2}\geq0$.}

{\bf Proof.} \ By Theorem 3.4, $W=Q^{T_2}$ as $W$ is optimal. If
there exists a positive operator $A$ such that $R(A)\subseteq R(Q)$
and $A^{T_2}\geq0$, then, by a well known result from operator
theory, there exists an operator  $T\in{\mathcal B}(H_1\otimes H_2)$
such that $A=QT$. It follows that $A^2=QTT^{\dagger}Q\leq tQ^2$,
where $t=\|T\|^2$. Thus, $A\leq\sqrt{t}Q$, which implies $Q-\lambda
A\geq0$ whenever  $0<\lambda<\frac{1}{\sqrt{t}}$. Thus we
 get $(Q-\lambda A)^{T_2}\in\mathcal{W}$. Now it follows from  Theorem 2.1
 (1)
that $(Q-\lambda A)^{T_2}$ is finer than $W$, a contradiction.
\hfill{$\square$}

{\bf Corollary 3.6.}\ {\it Let $H_1$, $H_2$ be complex Hilbert
spaces and $W\in{\mathcal W}(H_1\otimes H_2)$ be a decomposable
entanglement witness. If $W $ is optimal, then
$W^{T_2}\notin\mathcal{W}$.}

{\bf Proof.}  By Theorem 3.4, we know that $W=Q^{T_2}$ for some
$Q\geq0$. Therefore, $W^{T_2}=Q\geq0$. \hfill{$\square$}

For low dimensional systems, the optimal witnesses are easily
constructed. For example, the optimal witnesses for two qubits
(i.e., the $2\times2$ system) are of the form
$$W=|\psi\rangle\langle\psi|^{T_2},$$ where $|\psi\rangle$ is an
entangled state vector \cite{HP3}. In fact, an optimal witness
detecting the state $\rho$ can be constructed from the eigenvector
$|\psi\rangle$ of $\rho^{T_2}$ with negative eigenvalue $\lambda$ as
$W=|\psi\rangle\langle\psi|^{T_2}$ since ${\rm
Tr}(|\psi\rangle\langle\psi|^{T_2}\rho)={\rm
Tr}(|\psi\rangle\langle\psi|\rho^{T_2})=\lambda<0$ \cite{HP3}. This
method can be generalized to infinite-dimensional case but the
resulting witness may be not an optimal one.

\section{Witnesses without the finer relation between them}

Now we turn back to the question  when different entanglement
witnesses without ``finer'' relation between them can detect some
entangled states simultaneously. This question was studied in
\cite{WH} for finite-dimensional cases, there \cite[Theorem 4]{WH}
asserts that, in finite-dimensional systems, under the condition
${\rm Tr}(W_1)={\rm Tr}(W_2)$, if there exists no inclusion relation
between $\mathcal{D}_{W_1}$ and $\mathcal{D}_{W_2}$, then
$\mathcal{D}_{W_1}\cap \mathcal{D}_{W_2}\neq\emptyset$ if and only
if $W=\lambda W_1+(1-\lambda)W_2$ is not a positive operator for all
$0\leq\lambda\leq1$. We point out, though the result is true,
 the proof of \cite{WH} is not correct.

   Our attention is main focus on the
infinite-dimensional cases. We generalize the above result without
the assumption ``${\rm Tr}(W_1)={\rm Tr}(W_2)$'' and provide a proof
that valid  for both finite-dimensional systems and
infinite-dimensional systems.

 The following two lemmas are obvious.

{\bf Lemma 4.1.} {\it Let $H_1$, $H_2$ be  complex Hilbert spaces
and let $W_1$, $W_2\in\mathcal{W}(H_1\otimes H_2)$ with $W_1\prec
W_2$. If $W(a,b)=aW_1+bW_2$, $a$ and $b$ are positive numbers, then
$W_1\prec W(a,b)\prec W_2$.}

Particularly, if $W_1\prec W_2$, then all convex combinations of
them are still witnesses.

{\bf Lemma 4.2.} {\it Let $H_1$, $H_2$ be  complex Hilbert spaces.
For $W_1$, $W_2\in\mathcal{W}(H_1\otimes H_2)$, let
$W=aW_1+bW_2\neq0$ with $a\geq0$ and $b\geq 0$, then
$\mathcal{D}_W\subset\mathcal{D}_{W_1}\cup\mathcal{ D}_{W_2}$ and
$\mathcal{D}_{W_1}\cap\mathcal{D}_{W_2}\subset \mathcal{D}_W$.}

The following is our key lemma which is obtained for
finite-dimensional cases in \cite{WH} with a different and longer
proof.

{\bf Lemma 4.3.} {\it Let $H_1$, $H_2$ be  complex Hilbert spaces.
For $W, W_1, W_2\in\mathcal{W}(H_1\otimes H_2)$, if
$\mathcal{D}_{W_1}\cap\mathcal{D}_{W_2}=\emptyset$ and if
$\mathcal{D}_W\subset\mathcal{D}_{W_1}\cup\mathcal{ D}_{W_2}$, then
either $\mathcal{D}_W\subset\mathcal{D}_{W_1}$ or
$\mathcal{D}_W\subset\mathcal{D}_{W_2}$.}

{\bf Proof.} Assume, on the contrary, that both ${\mathcal
D}_{W_1}\cap{\mathcal D}_{W}$ and ${\mathcal D}_{W_2}\cap{\mathcal
D}_{W}$ are nonempty. Take $\rho_i\in{\mathcal D}_{W_i}\cap{\mathcal
D}_{W}$, $i=1,2$. Consider the segment
$[\rho_1,\rho_2]=\{\rho_t=(1-t)\rho_1+t\rho_2 : 0\leq t\leq 1\}$. As
${\mathcal D}_W$ is convex, we have
$$[\rho_1,\rho_2]\subseteq {\mathcal D}_W\subseteq {\mathcal
D}_{W_1}\cup{\mathcal D}_{W_2}. $$ Thus we get
$$[\rho_1,\rho_2]= ( {\mathcal
D}_{W_1}\cap [\rho_1,\rho_2])\cup({\mathcal
D}_{W_2}\cap[\rho_1,\rho_2]),$$ that is, $[\rho_1,\rho_2]$ is
divided into two convex parts. It follows that there is $0<t_0<1$
such that $\{\rho_t :0\leq  t<t_0\}\subseteq{\mathcal D}_{W_1}$,
$\{\rho_t : t_0<t\leq 1\}\subseteq{\mathcal D}_{W_2}$, and either
$\rho_{t_0}\in{\mathcal D}_{W_1}$ or $\rho_{t_0}\in{\mathcal
D}_{W_2}$. Assume that $\rho_{t_0}\in{\mathcal D}_{W_1}$; then ${\rm
Tr}(W_1\rho_{t_0})<0$. Thus, for sufficient small $\varepsilon>0$
with $t_0+\varepsilon\leq 1$, we have
$$0\leq {\rm Tr}(W_1\rho_{t_0+\varepsilon})={\rm
Tr}(W_1\rho_{t_0})+\varepsilon ({\rm Tr}(W_1\rho_2)-{\rm
Tr}(W_1\rho_1))<0,$$ a contradiction. Similarly,
$\rho_{t_0}\in{\mathcal D}_{W_2}$ leads to a contradiction, too.
This completes the proof.\hfill$\square$

Now we are ready to state and prove the main result in this section,
which asserts that two entanglement witnesses without ``finer''
relation between them can detect some entangled states
simultaneously if and only if all convex combinations of them are
entanglement witnesses.

{\bf Theorem 4.4.} {\it Let $H_1$, $H_2$ be  complex Hilbert spaces
and $W_1,
 W_2\in\mathcal{W}(H_1\otimes H_2)$. Then
$\mathcal{D}_{W_1}\cap\mathcal{D}_{W_2}=\emptyset$ if and only if
there exists $0<\lambda<1$ such that $W=\lambda
W_1+(1-\lambda)W_2\geq0$. }

By Lemma 4.1, Lemma 4.2 and Theorem 4.4, the following result is
immediate, which states that two witnesses can detect some entangled
states simultaneously if and only if their convex combination does
not break the witness.

{\bf Theorem 4.5.} {\it Let $H_1$, $H_2$ be  complex Hilbert spaces
and $W_1,
 W_2\in\mathcal{W}(H_1\otimes H_2)$. Then
$\mathcal{D}_{W_1}\cap\mathcal{D}_{W_2}\not=\emptyset$ if and only
if $W_\lambda =\lambda W_1+(1-\lambda)W_2\in {\mathcal W}(H_1\otimes
H_2)$ for all $0\leq \lambda\leq 1$.}

{\bf Proof of Theorem 4.4}. \ If  $W=\lambda W_1+(1-\lambda)W_2\geq
0$ for some $\lambda\in(0,1)$, then, by Lemma 4.2,
$\mathcal{D}_{W_1}\cap\mathcal{D}_{W_2}\subseteq {\mathcal
D}_W=\emptyset$.

Assume that $\mathcal{D}_{W_1}\cap\mathcal{D}_{W_2}=\emptyset$. Let
$W(\lambda)=\lambda W_1+(1-\lambda)W_2$, $0\leq\lambda\leq1$. Then,
by Lemma 4.3, for all $\lambda\in[0$, $1]$, we have
$$\mathcal{D}_{W(\lambda)}\subset\mathcal{D}_{W_1}, \
{\rm or}\ \mathcal{D}_{W(\lambda)}\subset\mathcal{D}_{W_2}.$$ When
$\lambda$ varies from 0 to 1 continuously,
$\mathcal{D}_{W(\lambda)}$ also varies from $\mathcal{D}_{W_2}$ to
$\mathcal{D}_{W_1}$ continuously. Taken $\lambda_0=\sup\{\lambda:
\mathcal{D}_{W(\lambda)}\subset\mathcal{D}_{W_2}\}$.

We claim that, if
$\mathcal{D}_{W(\lambda_0)}\subset\mathcal{D}_{W_2}$ then there
exist $0<\varepsilon<1-\lambda_0$ such that
$W(\lambda_0+\varepsilon)$ is a positive operator. Otherwise, if for
all $0<\varepsilon<1-\lambda_0$,
$\mathcal{D}_{W(\lambda_0+\varepsilon)}\neq\emptyset$, then we have
$$\mathcal{D}_{W(\lambda_0)}\subset\mathcal{D}_{W_2},\
\mathcal{D}_{W(\lambda_0+\varepsilon)}\subset\mathcal{D}_{W_1},$$
and for all $\rho\in \mathcal{D}_{W(\lambda_0)}$, we have $${\rm
Tr}(W(\lambda_0)\rho)<0, \ {\rm
Tr}(W(\lambda_0)\rho)+\varepsilon[{\rm Tr}(W_1\rho)-{\rm
Tr}(W_2\rho)]\geq0.$$ Noticing that ${\rm Tr}(W_1\rho))\geq0$ and
${\rm Tr}(W_2\rho))<0$, the second part of the last inequality is
positive, and $\varepsilon$ is an arbitrarily small positive number,
hence the last inequality is impossible. (We remark that there is a
mistake in the proof of \cite[Theorem 4]{WH} right here. In
\cite{WH}, the argument is `` for all $\rho\in
\mathcal{D}_{W(\lambda_0+\varepsilon)}$, we have
$${\rm Tr}(W(\lambda_0)\rho)\geq0, \ {\rm
Tr}(W(\lambda_0)\rho)+\varepsilon[{\rm Tr}(W_1\rho)-{\rm
Tr}(W_2\rho)]={\rm Tr}(W(\lambda_0+\varepsilon)\rho)<0.$$ Noticing
that ${\rm Tr}(W_1\rho)<0$ and ${\rm Tr}(W_2\rho)\geq0$, the second
part of the last inequality is negative, and $\varepsilon$ is an
arbitrarily small positive number, hence the last inequality is
impossible." However, ${\rm Tr}(W(\lambda_0)\rho)$ maybe equals 0
for all possible $\rho$ and the above argument is invalid.)

On the other hand, if $\mathcal{D}_{W(\lambda_0)}\subset\mathcal{
D}_{W_1}$ then there exist $0<\varepsilon<\lambda_0$ such that
$W(\lambda_0-\varepsilon)$ is a positive operator. Otherwise, if for
all $0<\varepsilon<\lambda_0$,
$\mathcal{D}_{W(\lambda_0-\varepsilon)}\neq\emptyset$, then we have
$$\mathcal{D}_{W(\lambda_0)}\subset\mathcal{ D}_{W_1},\
\mathcal{D}_{W(\lambda_0-\varepsilon)}\subset\mathcal{ D}_{W_2},$$
and for all $\rho\in\mathcal{D}_{W(\lambda_0)}$, we have $${\rm
Tr}(W(\lambda_0)\rho)<0, \ {\rm
Tr}(W(\lambda_0)\rho)+\varepsilon[{\rm Tr}(W_2\rho)-{\rm
Tr}(W_1\rho)]\geq0.$$ Noticing that ${\rm Tr}(W_2\rho))\geq0$ and
${\rm Tr}(W_1\rho))<0$, the second part of the last inequality is
positive, and $\varepsilon$ is an arbitrarily small positive number,
hence the last inequality is impossible (We remark that there is a
mistake similar to that pointed above in the proof of \cite[Theorem
4]{WH} right here, too.)

To sum up the discussion above, no matter
$\mathcal{D}_{W(\lambda_0)}\subset\mathcal{D}_{W_1}$ or
$\mathcal{D}_{W(\lambda_0)}\subset \mathcal{D}_{W_2}$ there exists
$\lambda\in[0$, $1]$ such that $W(\lambda)$ is a positive operator,
which completes the proof of the theorem.\hfill{$\square$}

In what follows, we generalize Theorem 4.4 and Theorem 4.5 by
allowing of finite many witnesses. The idea of the proof of the
statement (1) is similar to that in \cite{WH} for finite-dimensional
cases. Denote by ${\rm cov}(\Gamma)$ the convex hull of $\Gamma$,
that is, the convex set generalized by $\Gamma$.

{\bf Theorem 4.6.} {\it Let $H_1$, $H_2$ be  complex Hilbert spaces.
For a set of entanglement witnesses, $\Gamma=\{W_i:1\leq i\leq
n\}\subseteq \mathcal{W}(H_1\otimes H_2)$. Then}

(1) {\it $\mathcal{D}_\Gamma=\emptyset$ if and only if  ${\rm
cov}(\Gamma)$ contains some positive operators.}

(2) {\it $\mathcal{D}_\Gamma\not=\emptyset$ if and only if ${\rm
cov}(\Gamma)\subseteq{\mathcal W}(H_1\otimes H_2)$.}

{\bf Proof.} \ (1)  The sufficient part is clear. In fact, if
$W=\sum\limits_{i=1}^n\lambda_iW_i\geq0$ for some positive number
$\lambda_i$ with $\sum_i\lambda_i=1$, then
$\mathcal{D}_W=\emptyset$, which implies that
$\mathcal{D}_\Gamma=\emptyset$ since $\mathcal{D}_\Gamma\subseteq
\mathcal{D}_W$.

Conversely, if $\mathcal{D}_\Gamma=\emptyset$, we assume, without
loss of generality that, any subset of $\Gamma$ can detect some
entangled states simultaneously. If $n=2$, the theorem becomes
Theorem 4.4. Assume that the theorem holds for $k\leq n-1$. By
induction, we have to show that the theorem holds for $n$. Since the
method is the same,  we only need to show the case $n=3$. By
assumption, we have
$$\mathcal{D}_{W_1}\neq\emptyset, \ \mathcal{D}_{W_1}\cap
\mathcal{D}_{W_2}\neq\emptyset,\ \mathcal{D}_{W_1}\cap
\mathcal{D}_{W_3}\neq\emptyset,$$ but $$\mathcal{D}_{W_1}\cap
\mathcal{D}_{W_2}\cap \mathcal{D}_{W_3}=\emptyset,$$ namely,
$$(\mathcal{D}_{W_1}\cap \mathcal{D}_{W_2})\cap
(\mathcal{D}_{W_1}\cap \mathcal{D}_{W_3})=\emptyset.$$ Let
$$W(\lambda)=\lambda W_2+(1-\lambda)W_3,\ \lambda\in[0, 1],$$ then
$$\mathcal{D}_{W_1}\cap
\mathcal{D}_{W(\lambda)}\subset(\mathcal{D}_{W_1}\cap
\mathcal{D}_{W_2})\cup (\mathcal{D}_{W_1}\cap \mathcal{D}_{W_3}).$$
Since $\mathcal{D}_{W_1}\cap \mathcal{D}_{W_2}$ and
$\mathcal{D}_{W_1}\cap \mathcal{D}_{W_3}$ are disjoint, and
$\mathcal{D}_{W_1}\cap \mathcal{D}_{W(\lambda)}$ is convex, we know
that $\mathcal{D}_{W_1}\cap \mathcal{D}_{W(\lambda)}$ varies from
$\mathcal{D}_{W_1}\cap \mathcal{D}_{W_3}$ to $\mathcal{D}_{W_1}\cap
\mathcal{D}_{W_2}$ whenever $\lambda$ varies from 0 to 1. Using the
similar argument as that in the proof of Theorem 4.4, we can
conclude that there exists $0<\lambda_0<1$ such that
$$\mathcal{D}_{W_1}\cap \mathcal{D}_{W(\lambda_0)}=\emptyset.$$
Therefore, $$W=\mu W_1+(1-\mu)W(\lambda_0)=\mu W_1+(1-\mu)\lambda_0
W_2+(1-\mu)(1-\lambda_0)W_3\geq0$$ for some $\mu\in(0$, $1)$. By
induction on $n$, we complete the proof of (1).

(2) The ``only if'' part is obvious. To check the ``if'' part,
assume that ${\rm cov}(\Gamma)\subseteq{\mathcal W}(H_1\otimes
H_2)$. If, on the contrary, ${\mathcal D}_{\Gamma}=\emptyset$, then,
by the statement (1) just proved above, there exists $W\in{\rm
cov}(\Gamma)$ such that $W\geq 0$. It follows that
$W\not\in{\mathcal W}$, a contradiction. \hfill{$\square$}

By Theorem 4.6 it is clear that $W_1,\ldots , W_n\in{\mathcal
W}(H_1\otimes H_2)$ detect some entangled states simultaneously if
and only if all convex combinations of them are witnesses.

\section{Conclusion}

To sum up, in this paper, we answer the question under what
conditions different witnesses may detect some entangled states
simultaneously. Generally speaking, for  bipartite quantum systems,
finite many different witnesses can detect some entangled states
simultaneously if and only if their convex combinations are still
witnesses; they can not detect any entangled state simultaneously if
and only if the set of their convex combinations contains  a
positive operator. For two witnesses $W_1$ and $W_2$, some more can
be said:  (1) $W_1$ and $W_2$ can detect the same set of entangled
states if and only if they are linearly dependent; (2) $W_2$ can
detect more entangled states than
 that $W_1$ can if and only if $W_1$ is a linear combination of $W_2$ and
a positive operator. As an application of above results, we show
that a witness is optimal if and only if any negative permutation of
it will break the witness, that is, a witness $W$ is optimal if and
only if $W-D$ is not a witness for any positive operator $D$; $W$ is
decomposable optimal implies that $W$ is a partial transpose of some
positive operator.

Before the end, we would like to stress that our results holds for
both infinite-dimensional and finite-dimensional cases. Though some
of them are known for finite-dimensional systems under the
additional assumption ${\rm Tr}(W_1)={\rm Tr}(W_2)$, the proof of
our results are quite different.

\end{document}